# Multivariate data analysis: The French way

**Susan Holmes**[*,1]

*Stanford University*

**Abstract:** This paper presents exploratory techniques for multivariate data, many of them well known to French statisticians and ecologists, but few well understood in North American culture. We present the general framework of duality diagrams which encompasses discriminant analysis, correspondence analysis and principal components, and we show how this framework can be generalized to the regression of graphs on covariates.

## 1. Motivation

David Freedman is well known for his interest in multivariate projections [5] and his skepticism with regards to model-based multivariate inference, in particular in cases where the number of variables and observations are of the same order (see Freedman and Peters [12, 13]).

Brought up in a completely foreign culture, I would like to share an alien approach to some modern multivariate statistics that is not well known in North American statistical culture. I have written the paper *'the French way'* with theorems and abstract formulation in the beginning and examples in the latter sections; Americans are welcome to skip ahead to the motivating examples.

Some French statisticians, fed Bourbakist mathematics and category theory in the 60's and 70's as all mathematicians were in France at the time, suffered from abstraction envy. Having completely rejected the probabilistic enterprise as useless for practical reasons, they composed their own abstract framework for talking about data in a geometrical context. I will explain the framework known as the duality diagram developed by Cazes, Cailliez, Pagès, Escoufier and their followers. I will try to show how aspects of the general framework are still useful today and how much every idea from Benzecri's correspondence analysis to Escoufier's conjoint analysis has been rediscovered many times. Section 2.1 sets out the abstract picture. Sections 2.2-2.6 treat extensions of classical multivariate techniques: principal components analysis, instrumental variables, canonical correlation analysis, discriminant analysis, correspondence analysis from this unified view. Section 3 shows how the methods apply to the analysis of network data.

---

[*]Supported by NSF Grant DMS-0241246.

[1]Stanford University, Department of Statistics, Sequoia Hall, 390 Serra Mall, Stanford, CA 94305-4065, USA, e-mail: susan@stat.stanford.edu

*Keywords and phrases:* bootstrap, correspondence analysis, duality diagram, RV-coefficient, STATIS.

## 2. The duality diagram

Established by the French school of "Analyse des Données" in the early 1970's, this approach was only published in a few texts [1] and technical reports [9], none of which were translated into English. My Ph.D. advisor, Yves Escoufier [8, 10] publicized the method to biologists and ecologists, presenting a formulation based on his RV-coefficient that I will develop below. The first software implementation of duality based methods described here were done in LEAS (1984), a Pascal program written for Apple II computers. The most recent implementation is the **R** package ade-4 (see Appendix A for a review of various implementations of the methods described here).

### 2.1. Notation

The data are $p$ variables measured on $n$ observations. They are recorded in a matrix $X$ with $n$ rows (the observations) and $p$ columns (the variables). $D_n$ is an $n \times n$ matrix of weights on the "observations", which is most often diagonal. We will also use a "neighborhood" relation (thought of as a metric on the observations) defined by taking a symmetric definite positive matrix $Q$. For example, to standardize the variables $Q$ can be chosen as

$$Q = \begin{pmatrix} \frac{1}{\sigma_1^2} & 0 & 0 & 0 & ... \\ 0 & \frac{1}{\sigma_2^2} & 0 & 0 & ... \\ 0 & 0 & \frac{1}{\sigma_3^2} & 0 & ... \\ ... & ... & ... & 0 & \frac{1}{\sigma_p^2} \end{pmatrix}.$$

These three matrices form the essential "triple" $(X, Q, D)$ defining a multivariate data analysis. As the approach here is geometrical, it is important to see that $Q$ and $D$ define geometries or inner products in $\mathbb{R}^p$ and $\mathbb{R}^n$, respectively, through

$$x^t Q y = <x, y>_Q, \qquad x, y \in \mathbb{R}^p,$$
$$v^t D w = <v, w>_D, \qquad x, y \in \mathbb{R}^n.$$

From these definitions we see there is a close relation between this approach and kernel based methods, for more details see [24]. $Q$ can be seen as a linear function from $\mathbb{R}^p$ to $\mathbb{R}^{p*} = \mathcal{L}(\mathbb{R}^p)$, the space of scalar linear functions on $\mathbb{R}^p$. $D$ can be seen as a linear function from $R^n$ to $\mathbb{R}^{n*} = \mathcal{L}(\mathbb{R}^n)$. Escoufier[8] proposed to associate to a data set an operator from the space of observations $\mathbb{R}^p$ into the dual of the space of variables $\mathbb{R}^{n*}$. This is summarized in the following diagram [1] which is made commutative by defining $V$ and $W$ as $X^t D X$ and $X Q X^t$ respectively, (commutative just says that $VQ = X^t D X Q$ and $WD = X Q X^t D$).

We call $VQ$ the characterizing operator of the diagram.

$$\begin{array}{ccc} \mathbb{R}^{p*} & \xrightarrow{X} & \mathbb{R}^n \\ Q \uparrow \quad \downarrow V & D \downarrow \quad \uparrow W \\ \mathbb{R}^p & \xleftarrow{X^t} & \mathbb{R}^{n*} \end{array}$$

This is known as the duality diagram because knowledge of the eigendecomposition of $X^t DXQ = VQ$ leads to that of the dual operator $XQX^t D$. The main consequence is an easy transition between principal components and principal axes as we will see in the next section. The terms duality diagram or triple are often used interchangeably.

**Remarks.**

1. The duality diagram is equivalent to a triple of three matrices $(X, Q, D)$ such that $X$ is $n \times p$ and $Q$ and $D$ are symmetric matrices of the right size ($Q$ is $p \times p$ and D is $n \times n$). The operators defined as $XQX^t D = WD$ and $X^t DXQ = VQ$ are called the *c*haracteristic operators of the diagram [8]. We say an operator $O$ is $B$-symmetric if $< x, Oy >_B = < Ox, y >_B$, or equivalently $BO = O^t B$. In particular, $VQ$ is $Q$-symmetric and $WD$ is $D$-symmetric.
2. $V = X^t DX$ will be the variance-covariance matrix if $X$ is centered with regards to $D$ ($X'D\mathbf{1}_n = 0$) and $D$ is the diagonal matrix with all elements equal to $\frac{1}{n}$.
3. There is an important symmetry between the rows and columns of $X$ in the diagram, and one can imagine situations where the role of observation or variable is not uniquely defined. For instance in microarray studies the genes can be considered either as variables or observations. This makes sense in many contemporary situations which evade the more classical notion of $n$ observations seen as a random sample of a population. It is certainly not the case that the 30,000 probes are a sample of genes since these probes try to be an exhaustive set.

*2.1.1. Properties of the diagram*

Here are some of the properties that prove useful in various settings:

- Rank of the diagram: $X, X^t, VQ$ and $WD$ all have the same rank $r$, which will usually be smaller than both $n$ and $p$.
- For $Q$ and $D$ symmetric matrices, $VQ$ and $WD$ are diagonalisable and have the same eigenvalues. We denote them in decreasing order

$$\lambda_1 \geq \lambda_2 \geq \lambda_3 \geq \cdots \geq \lambda_r \geq 0 = \cdots = 0.$$

- Eigendecomposition of the diagram: $VQ$ is $Q$ symmetric, thus we can find $Z$ such that

(2.1) $$VQZ = Z\Lambda, \qquad Z^t QZ = \mathbb{I}_p,$$

where

$\Lambda = diag(\lambda_1, \lambda_2, \ldots, \lambda_r, 0, \ldots, 0)$ and $\mathbb{I}_p$ is the identity matrix in $\mathbb{R}^p$.

This generalized eigendecomposition of $VQ$ is often called the (generalized) PCA of the triple $(X, Q, D)$.

In practical computations, we start by finding the Cholesky decompositions of $Q$ and $D$, which exist as long as these matrices are symmetric and positive definite; call these $H^t H = Q$ and $K^t K = D$. Here $H$ and $K$ are upper triangular. Then we can use the singular value decomposition of $KXH^t$:

$$KXH^t = UST^t, \qquad \text{with } T^t T = \mathbb{I}_p, U^t U = \mathbb{I}_n, S \text{ diagonal},$$

to give us

$$X = K^{-1}UST^t(H^t)^{-1} = K^{-1}UST^t(H^{-1})^t \quad \text{and} \quad X^t = H^{-1}TSU^t(K^t)^{-1}.$$

Thus
$$HX^tDXH^t = TS^2T^t = T\Lambda T^t \text{ with } \Lambda = S^2$$

and finally we can see that $Z = H^{-1}T$ satisfies (2.1).

The renormalized columns of $Z$, $A = SZ$ are called the principal axes and satisfy:
$$A^tQA = \Lambda.$$

Similarly, we can define $L = K^{-1}U$ that satisfies

(2.2) $\quad WDL = L\Lambda, L^tDL = \mathbb{I}_n$, where $\Lambda = diag(\lambda_1, \lambda_2, \ldots, \lambda_r, 0, \ldots, 0)$.

$C = LS$ is usually called the matrix of principal components. It is normed so that
$$C^tDC = \Lambda.$$

When we impose that $C$ or $Z$ be of reduced rank $q < min(n, p)$, we take just their first $q$ columns, and have thus achieved what is known as the generalized PCA of rank $q$.

- Transition Formulæ: Of the four matrices $Z, A, L$ and $C$ we only have to compute one, all others are obtained by the transition formulæ provided by the duality property of the diagram:

$$XQZ = LS = C, \qquad X^tDL = ZS = A.$$

- The $Trace(VQ) = Trace(WD)$ is often called the inertia of the diagram (inertia in the sense of Huyghens' inertia formula for instance). The inertia with regards to a point $A$ of a cloud of $p_i$-weighted points being $\sum_{i=1}^{n} p_i d^2(x_i, a)$. When we look at ordinary PCA with $Q = \mathbb{I}_p$, $D = \frac{1}{n}\mathbb{I}_n$, and the variables are centered, the inertia is the sum of the variances of all the variables. If the variables are standardized ($Q$ is the diagonal matrix of inverse variances), then the inertia is the number of variables $p$.

### 2.2. Comparing two diagrams: the RV coefficient

Many problems can be rephrased in terms of comparison of two "duality diagrams" or put more simply, two characterizing operators, built from two "triples", usually with one of the triples being a response or having constraints imposed on it. We usually try to make one triple match the other in some optimal way.

To compare two symmetric operators, there is either a vector covariance $covV(O_1, O_2) = Tr(O_1^t O_2)$ or a vector correlation [8]

$$RV(O_1, O_2) = \frac{Tr(O_1^t O_2)}{\sqrt{Tr(O_1^t O_1)tr(O_2^t O_2)}}.$$

If we were in the special case of comparing two variables $X$ and $Y$ then the computation of the RV coefficient comparing the two triples $(X_{n\times 1}, 1, \frac{1}{n}I_n)$ and $(Y_{n\times 1}, 1, \frac{1}{n}I_n)$ would give the square of the correlation between the variables $RV = \rho^2$. Thus we see that in general the RV coefficient is an extension of the notion of correlation to the multivariate context.

Generalized PCA of rank $q$ of a $D$ centered matrix $X$ as defined above can be seen as providing best approximation $F$ in the RV-sense. To be more precise, we are looking for the matrix $F$ of rank $q$ which once inserted in a triple with the same weights on the observations $D$ and no weighting of the variables will maximizes the $RV$ coefficient between characterizing operators. Thus $F$ is the choice of matrix of rank $q < p$ that maximizes

$$RV\left(XQX^tD, FF^tD\right) = \frac{Tr\left(XQX^tDFF^tD\right)}{\sqrt{Tr\left(XQX^tD\right)^2 Tr\left(FF^tD\right)^2}}.$$

This maximum is attained where $F$ is chosen as the matrix combining the first $q$ eigenvectors of $XQX^tD$ normed so that $F^tDF = \Lambda_q$, the diagonal matrix where only the first $q$ eigenvalues are non zero. The maximum $RV$ is

$$RVmax = \sqrt{\frac{\sum_{i=1}^{q} \lambda_i^2}{\sum_{i=1}^{p} \lambda_i^2}}.$$

Of course, classical PCA has $D = \frac{1}{n}\mathbb{I}$, $Q = \mathbb{I}$, but the extra flexibility is often useful. We define the distance between triplets $(X, Q, D)$ and $(Z, P, D)$ where $Z$ is also $n \times p$, as the distance deduced from the RV inner product between operators $XQX^tD$ and $ZPZ^tD$. In fact, the reason the French like this scheme so much is that most multivariate linear methods can be reframed in these terms. We will give a few examples such as Principal Component Analysis (PCA in English, ACP in French), Correspondence Analysis (CA in English, AFC in French), Discriminant Analysis (LDA in English, AFD in French), PCA with regards to instrumental variable (PCAIV in English, ACPVI in French) and Canonical Correlation Analysis (CCA in English, AC in French).

### 2.3. Explaining one diagram by another

Principal Component Analysis with respect to Instrumental Variables was a technique developed by C. R. Rao [25] to find the best set of coefficients in the multivariate regression setting where the response is multivariate, given by a matrix $Y$. In terms of diagrams and RV coefficients, this problem can be rephrased as that of finding $M$ to associate to $X$ so that $(X, M, D)$ is *as close as possible* to $(Y, Q, D)$ in the RV sense.

The answer is provided by defining $M$ such that

$$YQY^tD = \lambda XMX^tD.$$

If this is possible then the two eigendecompositions of the triple give the same answers. We simplify notation by the following abbreviations:

$$X^tDX = S_{xx}, \qquad Y^tDY = S_{yy}, \qquad X^tDY = S_{xy}$$

and

$$R = S_{xx}^{-1}S_{xy}QS_{yx}S_{xx}^{-1}.$$

Then

$$\| YQY^tD - XMX^tD \|^2 = \| YQY^tD - XRX^tD \|^2 + \| XRX^tD - XMX^tD \|^2.$$

The first term on the right hand side does not depend on $M$, and the second term will be zero for the choice $M = R$.

If we add the extra constraint that we only allow ourselves a rank $q$ approximation, with $q < \min(\mathrm{rank}\,(X), \mathrm{rank}\,(Y))$, the optimal choice of a positive definite matrix $M$ is to take $M = RBB^t R$ where the columns of $B$ are the eigenvectors of $X^t DXR$ with:

$$B = \left(\frac{1}{\sqrt{\lambda_1}}\beta_1, \ldots, \frac{1}{\sqrt{\lambda_q}}\beta_q\right) \text{ such that } \begin{cases} X^t DXR\beta_k = \lambda_k \beta_k, \\ \beta_k^t R\beta_k = \lambda_k, k = 1, \ldots, q, \\ \lambda_1 > \lambda_2 > \cdots > \lambda_q. \end{cases}$$

The PCA with regards to instrumental variables of rank $q$ is equivalent to the PCA of rank $q$ of the triple $(X, R, D)$ where

$$R = S_{xx}^{-1} S_{xy} Q S_{yx} S_{xx}^{-1}.$$

### 2.4. One diagram to replace two diagrams

Canonical correlation analysis was introduced by Hotelling [18] to find the common structure in two sets of variables $X_1$ and $X_2$ measured on the same observations. This is equivalent to merging the two matrices columnwise to form a large matrix with $n$ rows and $p_1 + p_2$ columns and taking as the weighting of the variables the matrix defined by the two diagonal blocks $(X_1^t DX_1)^{-1}$ and $(X_2^t DX_2)^{-1}$

$$Q = \left(\begin{array}{c|c} (X_1^t DX_1)^{-1} & 0 \\ \hline 0 & (X_2^t DX_2)^{-1} \end{array}\right)$$

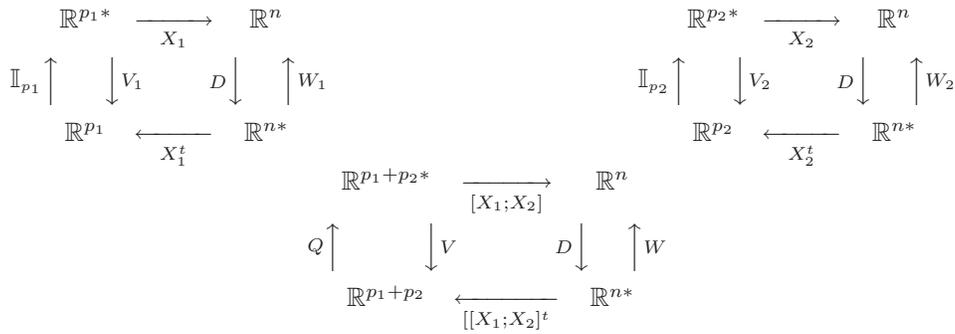

This analysis gives the same eigenvectors as the analysis of the triple $(X_2^t DX_1, (X_1^t DX_1)^{-1}, (X_2^t DX_2)^{-1})$, also known as the canonical correlation analysis of $X_1$ and $X_2$. These eigenvectors are known as the canonical variables.

### 2.5. Discriminant analysis

If we want to find linear combinations of the original variables $X_{n \times p}$ that characterize best the group structure of the points given by a zero/one group coding matrix $Y$, with as many columns as groups (call this number $g$), we can phrase the problem as a duality diagram. Suppose that the observations are given individual weights in the diagonal matrix $D$, and that the variables are centered with regards to these weights.

Let $A$ be the $g \times p$ matrix of group means in each of the $p$ variables. This satisfies $Y^t DX = \Delta_Y A$ where,

$$\Delta_Y = Y^t DY = diag(w_1, w_2, \ldots, w_g), \quad w_k = \sum_{i:y_{ik}=1} d_i.$$

The $w_k$'s are the group weights, as they are the sums of the weights as defined by $D$ for all the elements in that group. Call $T$ the matrix $T = X^t DX$, in the standard case with all diagonal elements of $D$ equal to $\frac{1}{n}$ this is just the standard variance-covariance, otherwise it is a generalization thereof. The generalized between group variance-covariance is $B = A^t \Delta_Y A$ and call the between group variance covariance the matrix $W = (X - YA)^t D(X - YA)$.

**Proposition 1.** (A generalized Huyghens' formula).

$$T = B + W.$$

*Proof.* Expanding $W$ gives

$$W = X^t DX - X^t DYA - A^t Y^t DX + A^t Y^t DYA$$
$$= T - A'\Delta_Y A - A'\Delta_Y A + A'\Delta_Y A$$
$$= T - B. \qquad \square$$

The duality diagram for linear discriminant analysis is

$$\begin{array}{ccc} \mathbb{R}^{p*} & \xrightarrow{A} & \mathbb{R}^g \\ T^{-1} \uparrow & \downarrow B \quad \Delta_Y \downarrow & \uparrow AT^{-1}A^t \\ \mathbb{R}^p & \xleftarrow{A^t} & \mathbb{R}^{g*} \end{array}$$

This corresponds to the triple $(A, T^{-1}, \Delta_Y)$, because

$$(X^t DY)\Delta_Y^{-1}(Y^t DX) = A^t \Delta_Y A$$

and gives equivalent results to the triple $(Y^t DX, T^{-1}, \Delta_Y^{-1})$.

The discriminating variables are the eigenvectors of the operator

$$A^t \Delta_Y A T^{-1}.$$

They can also be seen as the PCA with regards to instrumental variables of $(Y, \Delta_Y^{-1}, D)$ with regards to $(X, M, D)$.

### 2.6. Correspondence analysis

Correspondence analysis can be used to analyse several types of multivariate data. All involve some categorical variables. Here are some examples of the type of data that can be decomposed using this method:

- Contingency Tables (cross-tabulation of two categorical variables).
- Multiple Contingency Tables (cross-tabulation of several categorical variables).

- Binary tables obtained by cutting continuous variables into classes and then recoding both these variables and any extra categorical variables into 0/1 tables, 1 indicating presence in that class. So for instance a continuous variable cut into three classes will provide three new binary variables of which only one can take the value one for any given observation.

To first approximation, correspondence analysis can be understood as an extension of principal components analysis (PCA) where the variance in PCA is replaced by an *inertia* proportional to the $\chi^2$ distance of the table from independence. CA decomposes this measure of departure from independence along axes that are orthogonal according to a $\chi^2$ inner product. If we are comparing two categorical variables, the simplest possible model is that of independence in which case the counts in the table would obey approximately the margin products identity. For an $m \times p$ contingency table $N$ with $n = \sum_{i=1}^{m} \sum_{j=1}^{p} n_{ij}$ observations and associated to the frequency matrix $F = \frac{N}{n}$. Under independence, the approximation

$$n_{ij} \doteq \frac{n_{i\cdot}}{n} \frac{n_{\cdot j}}{n} n$$

can also be written: $N \doteq cr^t n$ where

$$r = \frac{1}{n} N 1_p$$

is the vector of row sums of $F$ and $c^t = \frac{1}{n} N' 1_m$ are the column sums. The departure from independence is measured by the $\chi^2$ statistic

$$\mathcal{X}^2 = \sum_{i,j} \left[ \frac{(n_{ij} - \frac{n_{i\cdot} n_{\cdot j}}{n^2} n)^2}{\frac{n_{i\cdot} n_{\cdot j}}{n^2} n} \right].$$

Under the usual validity assumptions that the cell counts $n_{ij}$ are not too small, this statistic is distributed as a $\chi^2$ with $(m-1)(p-1)$ degrees of freedom if the data are independent. If we do not reject independence, there is no more to be said about the table, no interaction of interest to analyse. There is in fact no 'multivariate' effect.

On the contrary if this statistic is large, we decompose it into one dimensional components.

Correspondence analysis is equivalent to the eigendecomposition of the triple $(X, Q, D)$ with

$$X = D_r^{-1} F D_c^{-1} - \mathbf{1}^t \mathbf{1}, \quad Q = D_c, \quad D = D_r,$$

$D_c = \text{diag}(c)$, $D_r = \text{diag}(r)$, $X' D_r \mathbf{1}_m = \mathbf{1}_p$, the average of each column is one.
Notes:

1. Consider the matrix $D_r^{-1} F D_c^{-1}$ and take the principal components with regards to the weights $D_r$ for the rows and $D_c$ for the columns.
   The recentered matrix $D_r^{-1} F D_c^{-1} - 1'_m 1_p$ has a generalized singular value decomposition

   $$D_r^{-1} F D_c^{-1} - 1'_m 1_p = U S V', \text{ with } U' D_r U = I_m, V' D_c V = I_p$$

   having total inertia:

   $$D_r (D_r^{-1} F D_c^{-1} - 1'_m 1_p)' D_c (D_r^{-1} F D_c^{-1} - 1'_m 1_p) = \frac{\mathcal{X}^2}{n}.$$

2. PCA of the row profiles $FD_r^{-1}$, taken with weight matrix $D_c$ and the metric $Q = D_c^{-1}$.
3. Notice that

$$\sum_i f_{i\cdot}(\frac{f_{ij}}{f_{i\cdot}f_{\cdot j}} - 1) = 0$$

and the row and columns profiles are centered

$$\sum_j f_{\cdot j}(\frac{f_{ij}}{f_{i\cdot}f_{\cdot j}} - 1) = 0$$

This method has been rediscovered many times, the most recently by Jon Kleinberg's in his method for analyzing Hubs and Authorities [19]. See Fouss, Saerens and Renders [11] for a detailed comparison.

In statistics the most commonplace use of Correspondence Analysis is in ordination or seriation, that is , the search for a hidden gradient in contingency tables. As an example we take data analyzed by Cox and Brandwood [4] and Diaconis [6], who wanted to seriate Plato's works using the proportion of sentence endings in a given book with a given stress pattern. The seven books studied here are Republic, Laws, Critias, Philebus, Sophist, Timœus. We use abbreviations of these names as our column labels in the data analysis below. The stress patterns use the last five syllables of every sentence and combine long or short syllables (abbreviated by - and U in the data below). Thus there are 32 possible stress patterns, and 32 rows in our contingency table.

We propose the use of correspondence analysis on the table of frequencies of sentence endings, for a detailed analysis see Charnomordic and Holmes [2].

The first 10 row profiles (as percentages) are as follows:

```
      Rep Laws Crit Phil Pol Soph Tim
UUUUU 1.1 2.4  3.3  2.5  1.7 2.8  2.4
-UUUU 1.6 3.8  2.0  2.8  2.5 3.6  3.9
U-UUU 1.7 1.9  2.0  2.1  3.1 3.4  6.0
UU-UU 1.9 2.6  1.3  2.6  2.6 2.6  1.8
UUU-U 2.1 3.0  6.7  4.0  3.3 2.4  3.4
UUUU- 2.0 3.8  4.0  4.8  2.9 2.5  3.5
--UUU 2.1 2.7  3.3  4.3  3.3 3.3  3.4
-U-UU 2.2 1.8  2.0  1.5  2.3 4.0  3.4
-UU-U 2.8 0.6  1.3  0.7  0.4 2.1  1.7
-UUU- 4.6 8.8  6.0  6.5  4.0 2.3  3.3
.......etc (there are 32 rows in all)
```

The eigenvalue decomposition (called the scree plot) of the chi-square distance matrix (see [2]) shows that two axes out of a possible 6 (the matrix is of rank 6) will provide a summary of 85% of the departure from independence. This suggests that a planar representation will provide a good visual summary of the data.

|   | Eigenvalue | inertia % | cumulative % |
|---|---|---|---|
| 1 | 0.09170 | 68.96 | 68.96 |
| 2 | 0.02120 | 15.94 | 84.90 |
| 3 | 0.00911 | 6.86 | 91.76 |
| 4 | 0.00603 | 4.53 | 96.29 |
| 5 | 0.00276 | 2.07 | 98.36 |
| 6 | 0.00217 | 1.64 | 100.00 |

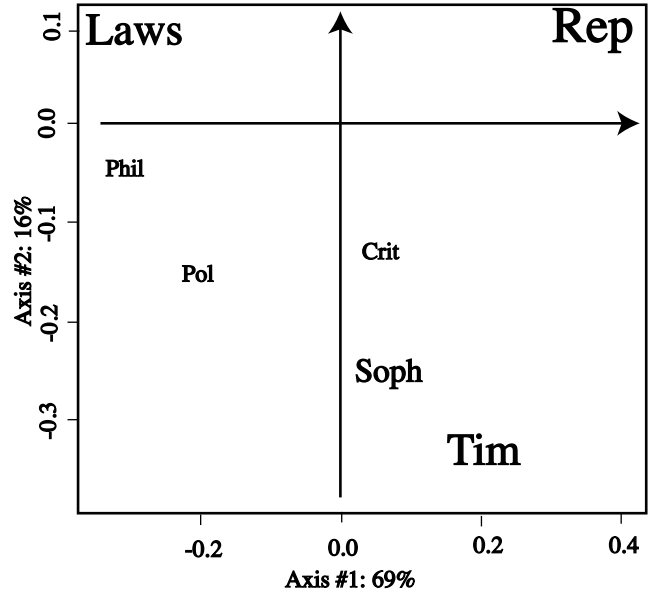

Fig 1. *Correspondence analysis of Plato's works.*

We can see from the plot that there is a seriation that in most cases follows a parabola or arch [16] from Laws on one extreme being the latest work and Republica being the earliest among those studied.

## 3. From discriminant analysis to networks

Consider a graph with vertices the members of a social group and edges if two members interact. We suppose each vertex comes with an observation vector $x_i$, and that each has the same weight $\frac{1}{n}$. In the extreme case of discriminant analysis, the graph is supposed to connect all the points of a group in a complete graph, and be disconnected between observations from different groups. Discriminant Analysis is just the explanation of this particular graph by linear combinations of variables. What we propose here is to extend this to more general graphs in a similar way. We will suppose all the observations are the nodes of the graph and each has the same weight $\frac{1}{n}$. The basic decomposition of the variance is written

$$\mathbf{cov}\,(x_j, x_k) = t_{jk} = \frac{1}{n}\sum_{i=1}^{n}(x_{ij} - \bar{x}_j)(x_{ik} - \bar{x}_k).$$

Call the group means,

$$\bar{\mathbf{x}}_{gj} = \frac{1}{n_g}\sum_{i\in G_g} x_{ij}, g = 1,\ldots,q,$$

$$\sum_{i\in G_g}(x_{ij} - \bar{\mathbf{x}}_{gj})(\bar{\mathbf{x}}_{gj} - \bar{x}_k) = (\bar{\mathbf{x}}_{gj} - \bar{x}_k)\sum_{i\in G_g}(x_{ij} - \bar{\mathbf{x}}_{gj}) = 0.$$

As in proposition 1, Huyghens' formula is $t_{jk} = w_{jk} + b_{jk}$, where

$$w_{jk} = \sum_{g=1}^{q} \sum_{i \in G_g} (x_{ij} - \bar{\mathbf{x}}_{gj})(x_{ik} - \bar{\mathbf{x}}_{gk}),$$

$$b_{jk} = \sum_{g=1}^{q} \frac{n_g}{n} (\bar{\mathbf{x}}_{gj} - \bar{x}_j)(\bar{\mathbf{x}}_{gk} - \bar{x}_k),$$

$$T = W + B.$$

As we showed above, linear discriminant analysis finds the linear combinations $a$ such that $\frac{a^t B a}{a^t T a}$ is maximized. This is equivalent to maximizing the quadratic form $a^t B a$ in $a$, subject to the constraint $a^t T a = 1$. As we saw above, the eigenvalue problem

$$Ba = \lambda T a \text{ or } T^{-1} Ba = \lambda a \text{ if } T^{-1} \text{ exists.}$$

provides $\lambda$ as needed. Then $a'Ba = \lambda a'Ta = \lambda$. We extend this to graphs by relaxing the group definition to partition the variation into local and global components.

### 3.1. Decomposing the variance into local and global components

Lebart was a pioneer in adapting the eigenvector decompositions to cater to spatial structure in the data [20, 21, 22]. We can again decompose the variance into parts, but this time the criteria for the decomposition is not defined by group membership as in LDA but by the neighborhood relation given by the spatial structure. We call the set of edges of the undirected neighborhood graph $E$. The usual elementwise definition of covariances is given by

$$\mathbf{cov}\,(x_j, x_k) = \frac{1}{n} \sum_{i=1}^{n} (x_{ij} - \bar{x}_j)(x_{ik} - \bar{x}_k) = \frac{1}{2n^2} \sum_{i=1}^{n} \sum_{i'=1}^{n} (x_{ij} - x_{i'j})(x_{ik} - x_{i'k}).$$

For the variances we have

$$\mathbf{var}(x_j) = \frac{1}{2n^2} \left\{ \sum_{(i,i') \in E} (x_{ij} - x_{i'j})^2 + \sum_{(i,i') \notin E} (x_{ij} - x_{i'j})^2 \right\}.$$

If we call $M$ the incidence matrix of the graph $m_{ij} = 1$ $(i,j) \in E$. The degree of vertex $i$ is $m_i = \sum_{i'=1}^{n} m_{ii'}$. We take the convention that there are no self loops. Then another way of writing the variance formula is

$$\mathbf{var}(x_j) = \frac{1}{2n^2} \left\{ \sum_{i=1}^{n} \sum_{i'=1}^{n} m_{ii'} (x_{ij} - x_{i'j})^2 + \sum_{(i,i') \notin E} (x_{ij} - x_{i'j})^2 \right\}.$$

Call local variance

$$\mathbf{var}_{loc}(x_j) = \frac{1}{2m} \sum_{i=1}^{n} \sum_{i'=1}^{n} m_{ii'} (x_{ij} - x_{i'j})^2$$

where $m = \sum_{i=1}^{n} \sum_{i'=1}^{n} m_{ii'}$. The total variance is the variance of the complete graph. Geary's ratio [14] is used to see whether the variable $x_j$ can be considered as independent of the graph structure. If the neighboring values of $x_j$ seem positively correlated then the local variance will only be an underestimate of the variance:

$$G = c(x_j) = \frac{\mathbf{var}_{loc}(x_j)}{\mathbf{var}(x_j)}.$$

Call $D$ the diagonal matrix with the total degrees of each node in the diagonal $D = diag(m_i)$.

For all variables taken together, $j = 1, \ldots, p$ note the local covariance matrix $\mathbf{V} = \frac{1}{2m} X^t (D - M) X$, if the graph is just made of disjoint groups of the same size. This is proportional to the $W$ within class variance-covariance matrix. The proportionality can be accomplished by taking the average of the sum of squares to the average of the neighboring nodes [23]. We can generalize the Geary index to account for irregular graphs coherently. In this case we weight each node by its degree. Then we can write the Geary ratio for any n-vector x as

$$c(x) = \frac{x^t(D-M)x}{x^t Dx}, \qquad D = \begin{pmatrix} m_1 & 0 & 0 & 0 \\ & m_2 & 0 & 0 \\ \vdots & & \ddots & \vdots \\ \ldots & 0 & 0 & m_n \end{pmatrix}.$$

We can ask for the coordinate(s) that are the most correlated to the graph structure, then if we want to minimize the Geary ratio, choose $x$ such that $c(x)$ is minimal. This is equivalent to minimizing $x^t(D - M)x$ under the constraint $x^t Dx = 1$. It can be solved by finding the smallest eigenvalue $\mu$ with eigenvector $x$ such that:

$$\begin{aligned} (D - M)x &= \mu Dx, \\ D^{-1}(D - M)x &= \mu x, \\ (1 - \mu)x &= D^{-1} Mx. \end{aligned}$$

This is exactly the defining equation of the correspondence analysis of the matrix M. This can be extended to as many coordinates as we like, in particular we can take the first 2 largest eigenvectors and provide the best planar representation of the graph in this way.

### 3.2. Regression of graphs on node covariates

The covariables measured on the nodes can be essential to understanding the fine structure of graphs. We call $\mathbf{X}$ the $n \times p$ matrix of measurements at the vertices of the graph; they may be a combination of both categorical variables (gene families, GO classes) and continuous measurements (expression scores). We can use the PCAIV method defined in Section 2 to the eigenvectors of the graph defined above. This provides a method that uses the covariates in $X$ to explain the graph. To be more precise, given a graph $(V, E)$ with adjacency matrix $M$, define the Laplacian

$$L = D^{-1}(M - I), \qquad D = diag(d_1, d_2, \ldots, d_n) \text{ diagonal matrix of degrees.}$$

Using the eigenanalysis of the graph, we can summarize the graph with a few variables, the first few relevant eigenvectors of $L$, these can then be regressed on the covariates using Principal Components with respect to Instrumental Variables [25] as defined above to find the linear combination of node covariates that explain the graph variables the best.

**Appendix A: Resources**

*A.1. Reading*

There are few references in English explaining the duality/operator point of view, apart from the already cited references of Escoufier [8, 10]. Fréderique Glaçon's PhD thesis [15] (in French) clearly lays out the duality principle before going on to explain its application to the conjoint analysis of several matrices, or data cubes. The interested reader fluent in French could also consult any one of several Masters level textbooks on the subject for many details and examples:

- Brigitte Escoffier and Jérôme Pagès [7] have a textbook with many examples, although their approach is geometric, they do not delve into the Duality Diagram, more than explaining on page 100 its use in transition formula between eigenbases of the different spaces.
- [22] is one of the broader books on multivariate analyses, making connections between modern uses of eigendecomposition techniques, clustering and segmentation. This book is unique in its chapter on stability and validation of results (without going as far as speaking of inference).
- Cailliez and Pagès [1] is hard to find, but was the first textbook completely based on the diagram approach, as was the case in the earlier literature they use transposed matrices.

*A.2. Software*

The methods described in this article are all available in the form of R packages which I recommend. The most complete package is `ade4` [3] which covers almost all the problems I mention except that of regressing graphs on covariates. However, a complete understanding of the duality diagram terminology and philosophy is necessary as these provide the building blocks for all the functions in the form of a class called `dudi` (this actually stands for duality diagram). One of the most important features in all the '`dudi.*`' functions is that when the argument `scannf` is at its default value `TRUE`, the first step imposed on the user is the perusal of the scree plot of eigenvalues. This can be very important, as choosing to retain 2 values by default before consulting the eigenvalues can lead to the main mistake that can be made when using these techniques: the separation of two close eigenvalues. When two eigenvalues are close the plane will be stable, but not each individual axis or principal component resulting in erroneous results if for instance the 2nd and 3rd eigenvalues were very close and the user chose to take 2 axes[17].

Another useful addition also comes from the ecological community and is called `vegan`. Here is a list of suggested functions from several packages:

- Principal Components Analysis (PCA) is available in `prcomp` and `princomp` in the standard package `stats` as `pca` in `vegan` and as `dudi.pca` in `ade4`.
- Two versions of PCAIV are available, one is called Redundancy Analysis (RDA) and is available as `rda` in `vegan` and `pcaiv` in `ade4`.
- Correspondence Analysis (CA) is available in `cca` in `vegan` and as `dudi.coa` in `ade4`.
- Discriminant analysis is available as `lda` in `stats`, as `discrimin` in `ade4`
- Canonical Correlation Analysis is available in `cancor` in `stats` (Beware `cca` in `ade4` is Canonical Correspondence Analysis).
- STATIS (Conjoint analysis of several tables) is available in `ade4`.

**Acknowledgments.** I would like to thank an anonymous referee for a very careful reading of the original version, Elizabeth Purdom for discussions about multivariate analysis and Yves Escoufier for reading this paper and teaching me much about Duality over the years. Persi Diaconis suggested looking at the Plato data in 1993 and has provided many enlightening discussions about the American way.